\documentclass[preprint2]{aastex6}  

\usepackage{graphicx}
\usepackage{textcomp}
\usepackage{wrapfig}
\usepackage{amsmath}
\usepackage{xspace}
\usepackage{xcolor}
\usepackage{blindtext, subfig}
\usepackage{dblfloatfix}

\newcommand{\Mathematica}{\texttt{Mathematica}\xspace}
\newcommand{\Python}{\texttt{Python}\xspace}

\shorttitle{Spatial distribution of fragments formed from tidally disrupted stars}

\shortauthors{Girma and Guillochon}

\begin{document}

   \title{The galactic distribution of fragments formed from tidally disrupted stars}

   \author{Eden Girma$^{\ast}$}
   \affiliation{Harvard College, Cambridge, MA 02138}
   \author{James Guillochon}
   \affiliation{
             Institute of Theory and Computation, Harvard-Smithsonian Center for Astrophysics, 60 Garden Street,
             Cambridge, MA 02138
             }
   \email{egirma@college.harvard.edu}
 
\begin{abstract}
Approximately once every $10^4$ years, a star passes close enough to the supermassive black hole Sgr~A* at the center of the Milky Way to be pulled apart by the black hole's tidal forces. The star is then ``spaghettified'' into a long stream of matter, with approximately one half being bound to Sgr~A* and the other half unbound. Within this stream, the local self-gravity dominates the tidal field of Sgr~A*, which at minimum restricts the stream to a small finite width. As the stream cools from adiabatic expansion and begins to recombine, the residual self-gravity allows for planetary-mass fragments to form along the length of the stream; these fragments are then shot out into the galaxy at range of velocities, with the fastest moving at $\sim$~10\%~$c$. We determine the phase space distributions of these fragments for a realistic ensemble of stellar disruptions, along with the local density of fragments in the solar neighborhood. We find that $\sim 10^7$ fragments produced by Sgr~A* accumulate within the Milky Way over its lifetime, that there are $\sim 10^7$ fragments that lie within 1~Mpc of the Milky Way originating from other galaxies, and that the nearest fragment to our Sun is on average 500~pc distant.
\end{abstract}
   


\section{Introduction}
Stars orbiting the supermassive black hole (SMBH) at the center of our galaxy have the potential to pass too close and be disrupted by the SMBH's overwhelming gravitational force. This tidal disruption event (TDE) is dependent on the star passing within a distance known as the tidal radius, determined by the mass of the SMBH ($M_{\rm BH}$) and the stellar mass and radius ($M_{\star}$, $R_{\star}$),
\begin{equation}
r_{t} \simeq R_{\star} \left(\frac{M_{\rm BH}}{M_{\star}}\right)^{\frac{1}{3}}\label{eq:rt}
\end{equation} at which the SMBH's gravity overpowers the star's self-gravity \citep{Rees:1988a}. A full or partially disrupted star is characterized by the impact parameter $\beta \equiv r_{\rm t}/r_{\rm p}$, where $r_{\rm p}$ is the star's distance to the SMBH at pericenter, with stars losing anywhere from a tiny fraction of their mass for $\beta \ll 1$ and being completely destroyed for $\beta \gg 1$ \citep{Guillochon:2013a}.

Hydrodynamical simulations of this process have demonstrated that the local self-gravity of the tidal stream dominates the tidal gravity of the hole \citep{Kochanek:1994a,Guillochon:2014a}. The self-gravity within the stream can result in the formation of fragments along the stream that are then launched out into the galaxy with a range of binding energies \citep{Coughlin:2015a}.  This fragmentation process poses various questions regarding the evolution of these objects. In this paper, we seek to answer: (1) How many fragments in total are produced by stellar disruptions over our own galaxy's life time, (2) what is the final spatial distribution of the fragments produced from TDEs originating at the galactic center, and (3) how near is the closest fragment to our sun?

Our simulation of these fragments' motion consists of an initialization package written in \Python and an N-body integrator based in \Mathematica. In Section~\ref{sec:fragmentation}, we present an analytic analysis of the environmental assumptions underlying our simulation, and in Section~\ref{sec:methods} describe in more detail the construction and steps taken with our \Python/\Mathematica code. Section~\ref{sec:results} presents the results of our simulation and the analysis. We conclude in Section~\ref{sec:discussion} with a discussion of our results' implications  and additional questions to be posed regarding fragment observability and the existence of fragments produced by TDEs in nearby galaxies.

\section{Fragmentation}\label{sec:fragmentation}

\citet{Coughlin:2015a} had found that $\gamma = 5/3$ debris streams were marginally stable, with the fragmentation occurring at later times for higher resolutions. This suggests that numerical perturbations were likely the seeds for the fragments formed in those simulations. While these particular perturbations were likely spurious, density perturbations are present in real stars (from e.g. convection), and thus the fragmentation effect that \citeauthor{Coughlin:2015a} observed is likely to be realized in nature. An important contributor to fragmentation processes in tidal streams is cooling by recombination \citep{Roos:1992a,Kasen:2010a,Guillochon:2014b}. Once cooling by recombination begins, perturbations within the stream can grow non-linearly, as the cooling rate is highly sensitive to temperature \citep{Sutherland:1993a}.

For the purposes of our simulation, we work under the assumption that fragments begin to form once the disrupted stream cools adiabatically to temperatures low enough for hydrogen recombination, $T_{\rm f}=5 \times 10^3$ K, at which point the cooling rate rapidly increases and fragmentation is likely to begin. The initial temperature $T_{\rm i}$ at the core of the star can be calculated with the stellar mass and radius, using the microscopic ideal gas law $P = n_{\rm i} k_{\rm b} T_{\rm i} $ where $n_i = \rho_{\star} / \mu m_{\rm p}$ is the initial number density calculated from stellar density $\rho_{\star}$, the mean molecular weight of hydrogen $\mu=0.5$, and the mass of a proton $m_{\rm p}$, 
\begin{align}
T_{\rm i} &= \frac{P}{n_{\rm i} k_{\rm b}} \nonumber\\
&= \left[\frac{1}{4 \pi R_{\star}^2}\left(\frac{GM_{\star}^2}{2R_{\star}}\right)\right] \left(\frac{\left(\frac{4}{3}\pi R_{\star}^3\right)\mu m_{\rm p}}{M_{\star}k_{\rm b}}\right)
\end{align} 
Assuming the gas is described as a $\gamma = 5/3$ polytrope, the change in volume is proportional to $T^{3/2}$, which yields an expansion factor $\alpha = \left(T_{\rm f}/T_{\rm i}\right)^{3/2}$. The final gas density when fragmentation occurs given a value $\beta$ is then 
\begin{equation}
n_{\rm f} = \frac{0.5 n_{\rm i} \mathcal{C}(\beta)}{\alpha},\label{eq:nf}
\end{equation} 
where $\mathcal{C}(\beta)$ is the fraction of stellar mass removed by the disruption, as defined in \citet{Guillochon:2013a,Guillochon:2015b}. The mass of all fragments are assumed to be equal to the Jeans mass,
\begin{align}
M_{\rm frag} &= \frac{\pi}{6} \frac{c_{\rm s}^3}{G^{3/2} \rho^{1/2}}\label{eq:mfrag}\\
&= 1.74\times 10^{-8} M_J c_{\rm s}^3 n_{\rm f}^{-1/2}\nonumber,
\end{align} 
where $c_{\rm s} = \sqrt{\gamma k_{\rm b} T_{\rm f}/ \mu m_{\rm p}}$ is the sound speed, with $\gamma = \frac{5}{3}$ for a gas-pressure dominated polytropic fluid, and $\rho = \rho_\ast / \alpha$.

The number of fragments can then be calculated by dividing the fraction of stellar mass removed in the disruption ($M_{\rm frag} = M_{\star} \mathcal{C}(\beta)$ given a specific $\beta$ value) by the mass of the fragment as determined in Equations (3-4),
\begin{equation}
N_{\rm frag}= \frac{0.5 M_{\star} \mathcal{C}(\beta)}{M_{\rm frag}}.
\end{equation}

\section{Methods}\label{sec:methods}

Our simulation is first initialized through a package written in \Python, which inputs the user-driven variables of number of stars disrupted\footnote{\url{https://github.com/edengirma/TDEfragModel/tree/master/fragSimSetup}}. The code randomly draws the parameters that define the star being disrupted (e.g. stellar mass, radius, tidal radius), and the disruption itself ($\beta$, the number of fragments produced, the specific binding energy spread). It then calculates, using these parameters, an initial position and velocity vector for each fragment. These values are written to an intermediate {\tt JSON} file and used as starting positions for a integrator written within a \Mathematica notebook. This integrator outputs as solutions for each fragment an interpolation function describing the evolution of $x$, $y$, and $z$ positions over the integrated time.

\subsection{Initial Conditions}
The mass of Sgr~A* is set to $M_{\rm BH} = 4 \times 10^6 M_{\odot}$. For each star, the stellar mass is randomly drawn over the interval $[0.1M_{\odot}, 100M_{\odot}]$ from a Salpeter initial mass function \citep{Salpeter:1955a}. We approximate the stellar radius as $R_{\star}\propto M_{\star}^{0.8}$ \citep{Kippenhahn:1990a} and use the tidal radius definition from Equation (\ref{eq:rt}). To calculate the number of fragments produced in a given simulated TDE, stars are assumed to be deposited near the SMBH via pinhole scattering \citep{Lightman:1977a}, and thus $\beta$ is drawn over the interval $[0.5,2.5]$ from a $\beta^{-2}$ probability distribution. The total mass of the stream is then calculated as a function of $\beta$ and used to determine $N_{\rm frag}$ given the stellar mass (Equation (\ref{eq:nf})).

\begin{figure}[t]
  \centering
  \includegraphics[width=\linewidth]{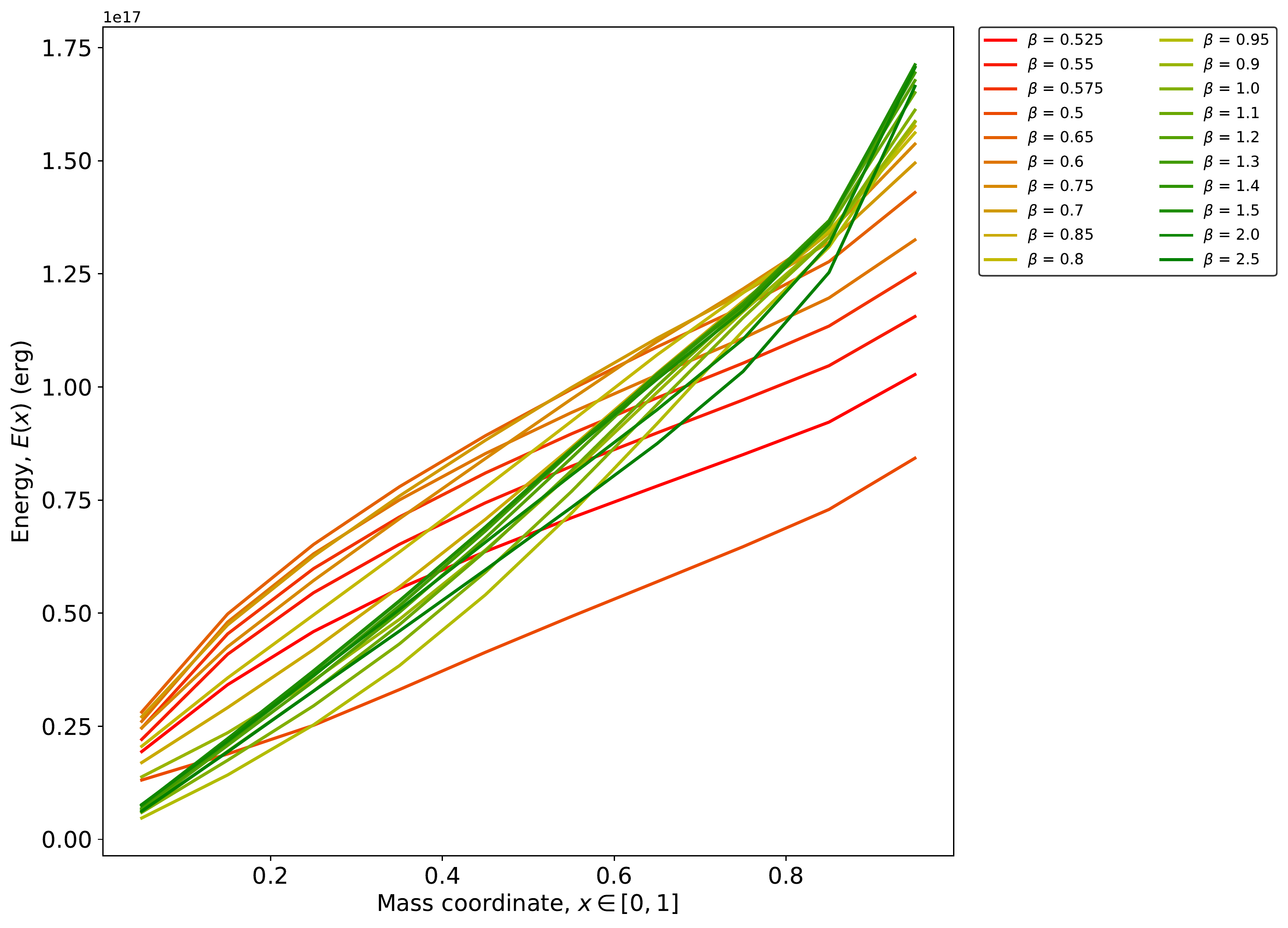}
  \caption{Interpolated binding energy distributions for different TDE impact parameters $\beta$, where $\beta$ ranges from a grazing partial encounter to a deep full disruption.}\label{fig:edist}
\end{figure}

The position vector $\vec{r}_{\star}$ of the star is set using random sphere point picking at a distance $r_{\rm p}$ away from the SMBH, which we assume lies at the exact center of the galaxy, and the direction of each fragment's velocity vector is randomly determined on a plane perpendicular to $\vec{r_{\star}}$ using a rotation angle $\phi \in [0,2\pi)$. The initial position of each fragment $\vec{r}_i$ corresponds to a radial distance from the SMBH in the range $r_{\rm p} \leq r_i \leq r_{\rm p} + R_{\star}$, with each distance corresponding to an orbital energy $E = G M_{\rm h} \delta r / r_{\rm p}^2$ that lies between zero (a parabolic orbit) and $E_{\max} \simeq G M_{\rm h} R_\star / r_{\rm p}^2$. As the star's original radial mass profile $M(r)$ is largely preserved in the disruption given its nearly-homologous expansion \citep{Coughlin:2016c}, we can determine $E$ (and thus $r_i$) for each fragment by partitioning $E$ based upon $M(r)$ into equal chunks with mass $M_{\rm frag}$ (Equation (\ref{eq:mfrag})). In Figure~\ref{fig:edist}, we show the resulting $E(x)$ for the unbound debris stream for various values of $\beta$ \citep[taken from][]{Guillochon:2013a}, where $x \equiv M(r)/M(R_\star) \in [0,1]$ is the mass coordinate. Assuming all fragments are the same mass, the energy of each fragment can be determined by partitioning $E(x)$ in equal increments $\delta x$, where $\delta x = 1 / N_{\rm frag}$. This binding energy is then multiplied by an energy scale, and the scaled energy is used to calculate total velocity of the fragment,
\begin{equation}
v_{\rm frag} = v_{\infty} + v_{\rm h} = \sqrt{2E} + \frac{2GM_{\rm BH}}{r_{\rm t}}.
\end{equation}

The above process is repeated $N$ times to simulate an ensemble of disruptions. The outputs are then passed to a {\tt Mathematica} notebook that handles the integration of the fragments' trajectories, which we describe in the next subsection.

\subsection{Evaluating Fragment Orbits} 
Given the initial position and velocity vectors for a fragment, we solve for the evolution of a fragment's position $\vec{r}_{\rm frag}(t)$ through the second order differential force equation $m_{\rm frag} \partial^{2} (\vec{r}_{\rm frag}(t)) /\partial t^{2} = \vec{F}$ at each time step $0 \leq t \leq \tau_{\rm H}$. The forces $\vec{F}$ that a fragment experiences include the gravitational attraction of Sgr~A* and forces derived from the gravitational potential of the Milky Way, as described in \cite{Kenyon:2014a},
\begin{align}
	F_i &= (F_{\rm BH})_i+(F_{\rm b})_i+(F_{\rm c})_i+(F_{\rm d})_i+(F_{\rm h})_i\\
\intertext{where $i=1,2,3$ indicates the $x$, $y$, and $z$ component of $\vec{F}$, and}
	(F_{\rm BH})_i &= \frac{-GM_{\rm BH}\vec{r}_i}{r^3}\\
	(F_{\rm b})_i &= \frac{-GM_b\vec{r}_i}{r^2(r_b+r)}\\
	(F_{\rm c})_i &= \frac{-2GM_{\rm BH}\vec{r}_i}{\max(r_c,r) r^2}\\
	(F_{\rm d})_i &= \frac{-GM_{\rm d}\vec{r}_i}{\left(x^2+y^2+\left[a_{\rm d}+\sqrt{z^2+b_{\rm d}^2}\right]^2\right)^{3/2}}\\
	(F_{\rm h})_i &= -GM_{\rm h}\vec{r}_i \left[\frac{\ln(1+\frac{r}{r_{\rm h}})}{r^3} - \frac{1}{r^2(r+r_{\rm h})}\right]
\end{align} are the forces due to Sgr~A*, the galaxy bulge, nuclear cluster, disk, and halo respectively. Numerical constants are set as defined in \citeauthor{Kenyon:2014a}; for the bulge, disk, and halo, $M_{\rm b} = 3.76 \times 10^9 M_{\odot}$, $M_{\rm d} = 6 \times 10^{10} M_{\odot}$, and $M_{\rm h} = 10^{12} M_{\odot}$. The radius of the halo and bulge are $r_{\rm h} = 20$ kpc and $r_{\rm b} = 0.1$ kpc. The parameters $a_{\rm d} = 2.75$ kpc and $b_{\rm d} = 0.3$ kpc are set such that the disk potential matches a circular velocity of $235$ km s$^{-1}$ at the position of the Sun.

Following the trajectory of a fragment from the galactic center to outskirts of the galaxy involves resolving forces on a wide variety of scales, from $\sim$~AU to $\sim$~Mpc (11 orders of magnitude). Over this distance different components will dominate the force that pulls the fragment back towards the galaxy: first the SMBH, then the nuclear cluster, then the galaxy's bulge, disk, and halo. We experimented with a few \Python-based integrators that are publicly available but were unable to find any that were able to handle the wide range of force scales within the same package.

\Mathematica, with its arbitrary-precision computations, possesses a powerful capacity to perform integrations even if the scales change rapidly over the course of a computation. Namely, it can calculate arbitrary order solutions which allow for accurate resolution on a range of scales, i.e. solutions that model both short-range distances when the force of the SMBH overpowers all other contributions, and long-range distances when the forces of the disk, bulge, and halo dominate. The integrator is used to evolve the positions and velocities of each fragment for each disruption, yielding a full temporal history of a particle's position. We halt the integration once a particle is determined to be bound to the galaxy (i.e. it has reached apoapse), or once $\tau_{\rm H} = 10^{10}~{\rm yr}$ have elapsed. Our solutions for each fragment's position are in the form interpolated functions, with varying domains of $t$ depending on the integration time.

\section{Results}\label{sec:results}

A set of $N_{\rm TDE} = 10^{3}$ TDEs were simulated, with the number of fragments per disruption determined from the randomly drawn $M_{\star}$ and $\beta$ as described in Section~\ref{sec:fragmentation}. A total of 8,923 fragments were found to be bound (8.9 on average per disruption) and 188,112 unbound (190 per disruption). Assuming a TDE rate $\Gamma_{\rm TDE} = 10^{-4}~{\rm yr}^{-1}$, this yields $9 \times 10^{6}$ fragments bound to the galaxy over $\tau_{\rm H}$. We found that 47\% of the bound fragments had an apoapsis within 0.1~pc, and 88\% of bound fragments had an apoapsis less than or equal to 100~pc. Thus, the vast majority of bound fragments end up closely bound to the immediate vicinity of the SMBH. A sizable drop in fragment count occurred beyond a kpc, with only 255 total fragments possessing an apoapsis within the range 1~--~10~kpc. However, restricting to ranges between a kpc and an Mpc, we find a small bump in fragment count: 4.0\% of fragments have an apoapsis in the range 10~--~100~kpc, while only 2.9\% and 1.9\% lie within the ranges 1~--~10~kpc and 100~kpc~--~1~Mpc, respectively. The fraction of bound fragments whose apoapsis lies within 1~kpc of the sun is small, 0.35\%.

The majority of the unbound fragments (78\%) traveled a distance 10~--~100~Mpc from the galactic center, as expected for objects moving at a few thousand km~s$^{-1}$ for a Hubble time. The maximum distance traveled by an unbound fragment is 130~Mpc, with 22\% of unbound fragments traveling at least 100~Mpc from the galaxy. Only 93 unbound fragments (0.86\%) traveled a distance less than 10~Mpc from the galactic center.

\begin{figure}[t]
	\centering
    \subfloat[Position histogram of fragments]{
    	\includegraphics[width=\linewidth]{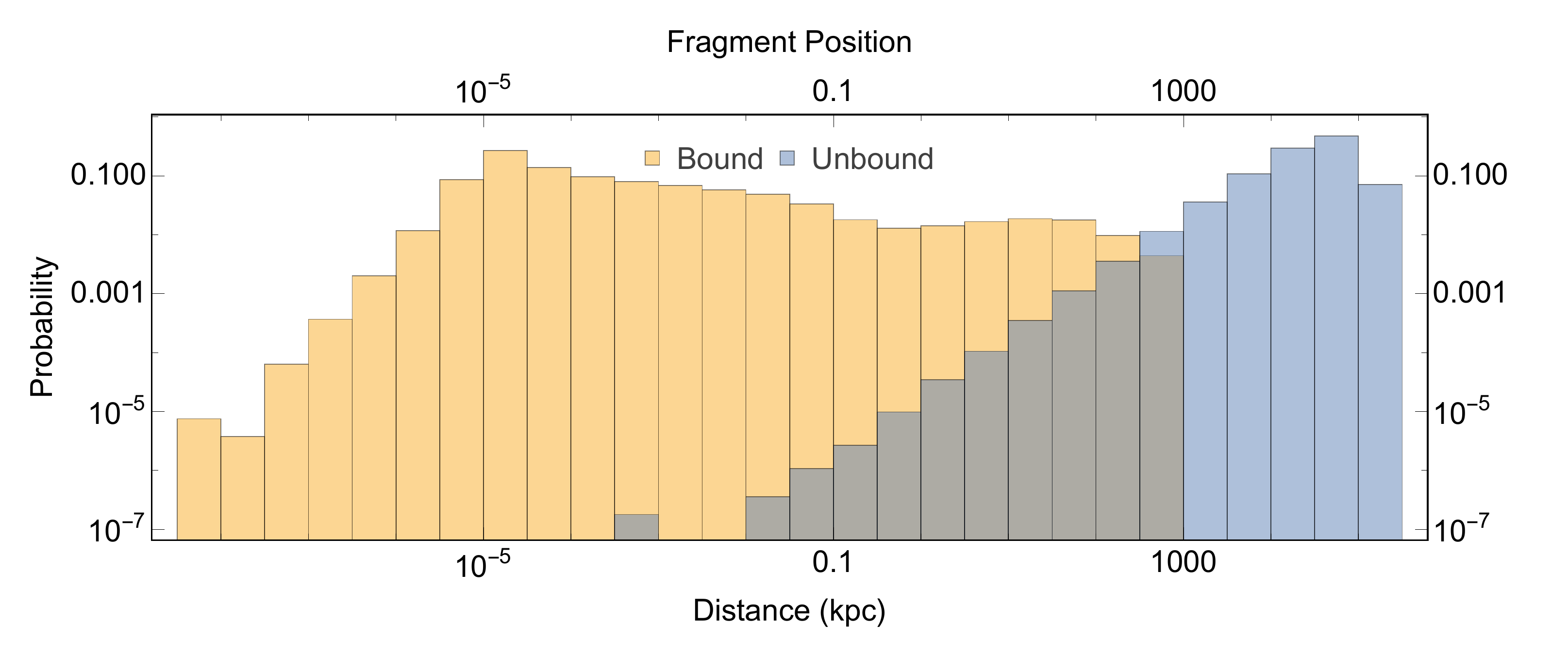}
    }\\*[3mm]
    \subfloat[Velocity histogram of fragments]{
    	\includegraphics[width=\linewidth]{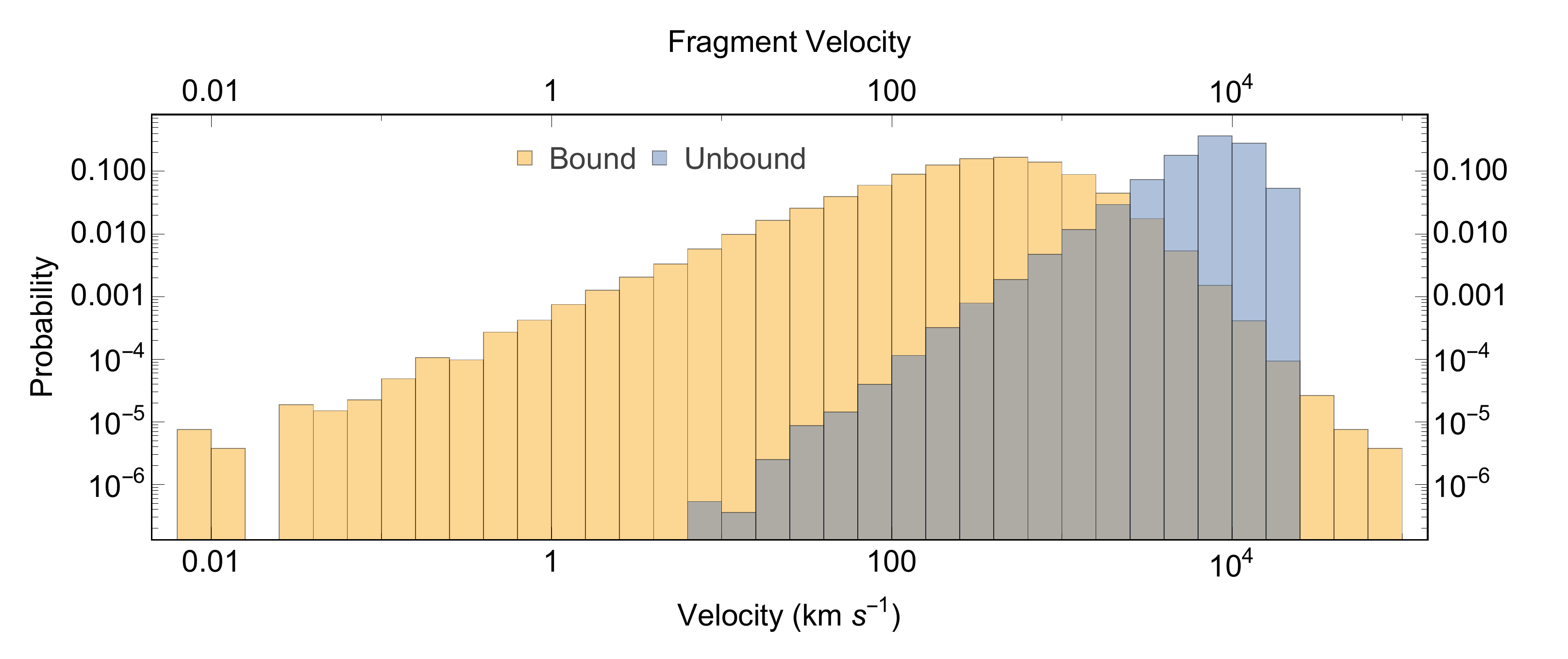}
    }
    \caption{Histograms of ${\cal N}_{\rm frag}$ fragment realizations of position and velocity. Both histograms are constructed by drawing positions and velocities from 30 random times in each fragment's orbit. Position measured in kpc and velocity in km~s$^{-1}$ are binned logarithmically.}\label{fig:histograms}
\end{figure}

To generate smooth distributions of fragment positions that represent the potentially observable population, we drew 30 random times for each fragment over a single orbital period for the bound fragments or $\tau_{\rm H}$ for the unbound, yielding approximately ${\cal N}_{\rm frag} = 6 \times 10^{6}$ fragment positions. The distance from the galactic center and the velocities of the bound and unbound fragments are shown in Figure~\ref{fig:histograms}. The resulting data, as outlined below, gives us a better sense of the resulting phase space distribution of fragments produced through the process of tidal disruption in our galaxy.

A vast majority of the potentially observable population's bound fragments (89\%) are concentrated within 100~pc of the galactic center. In comparison to the unbound fragments, bound fragments move significantly slower with an average speed of 580 km~s$^{-1}$; the average velocity of unbound fragments is $\sim 8 \times 10^3$ km~s$^{-1}$ ($0.03 c$), with 33\% moving at greater than $10^4$ km~s$^{-1}$. The maximum velocity of an unbound fragment was found to be 24,000 km~s$^{-1}$, or $0.08 c$, comparable to the fastest hypervelocity stars produced by black hole-black hole mergers \citep{Guillochon:2015c}.

\begin{figure}[t]
	\centering
    \subfloat[Spatial density plot, cartesian $x$ and $z$ coordinates.]{
    	\includegraphics[width=\linewidth]{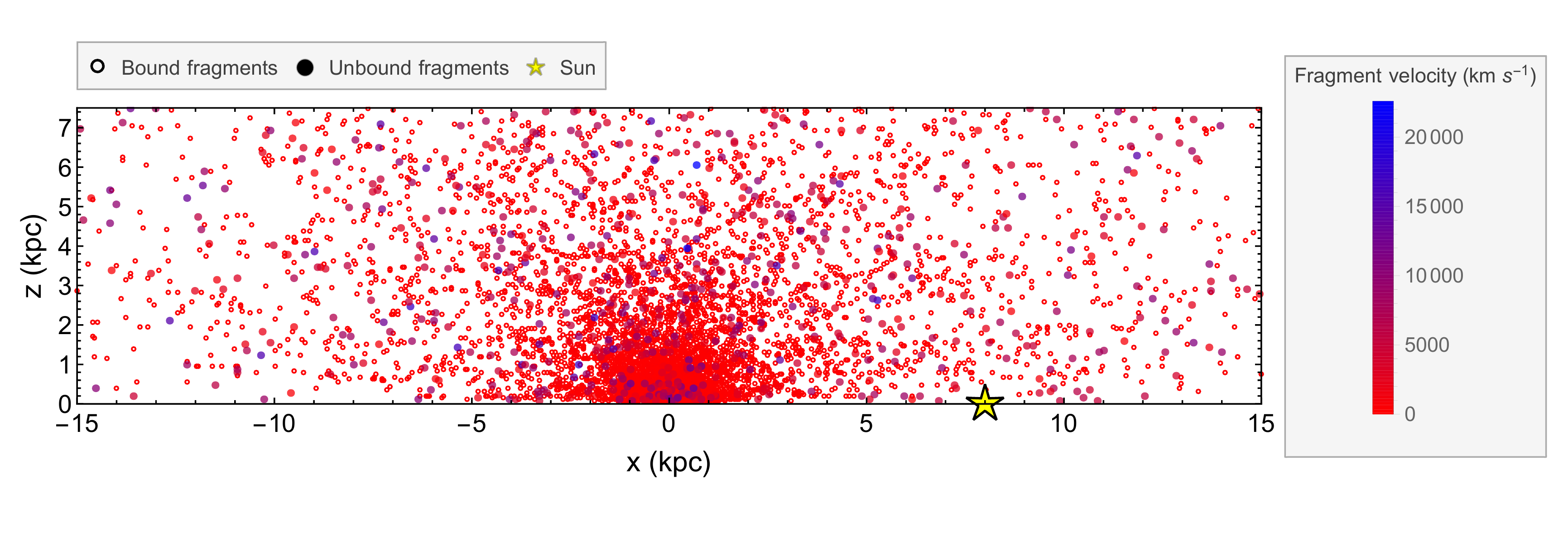}
    }\\*[3mm]
    \subfloat[Spatial density plot, cylindrical $s$ and $|z|$ coordinates.]{
    	\includegraphics[width=\linewidth]{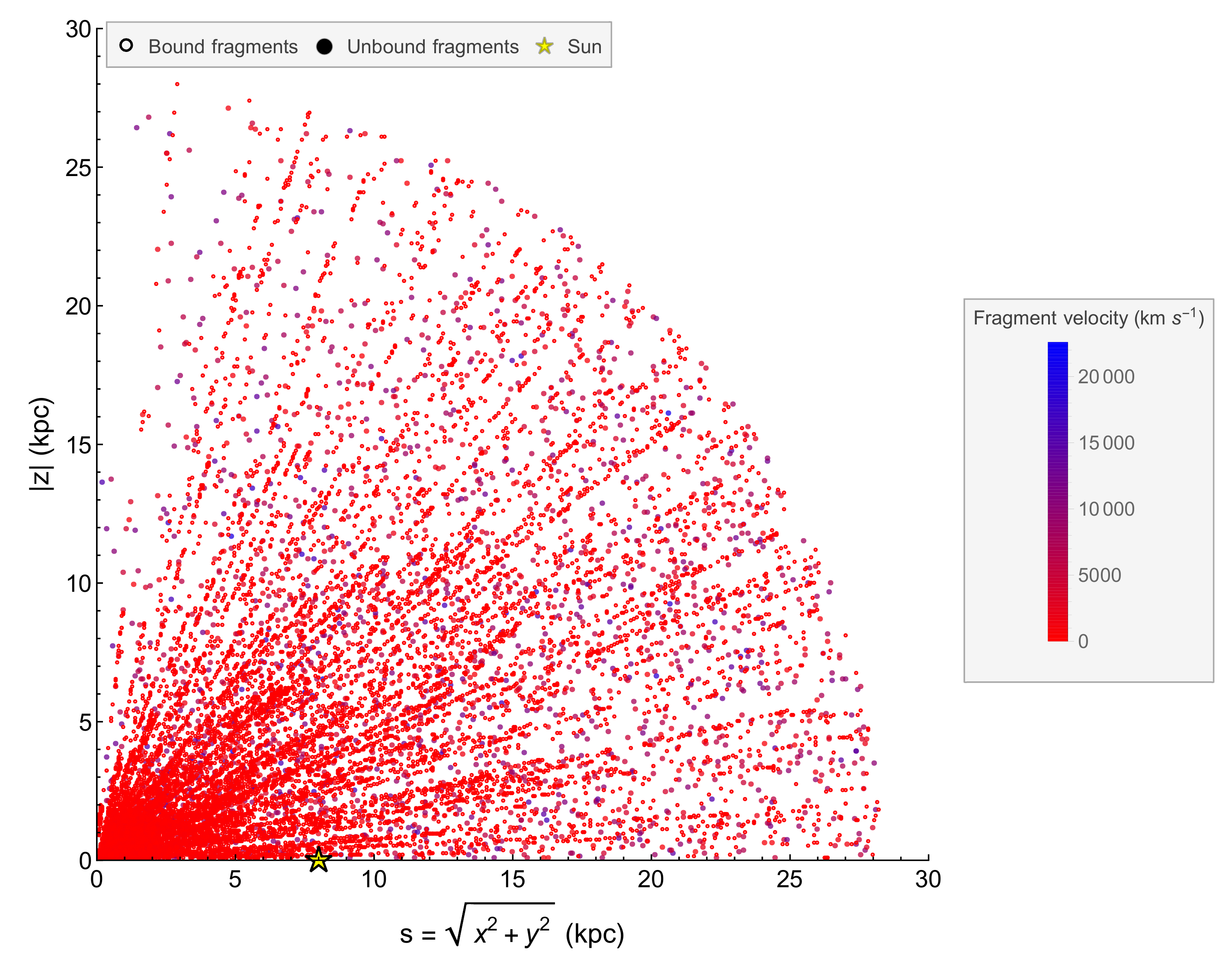}
    }
    \caption{
    Spatial density plots of the potentially observable fragments within the Milky Way. Bound and unbound fragments are differentiated as unfilled and filled circles, respectively. Each point is color-coded with its velocity at that position. Note that only a few unbound fragments, which were traveling at much higher speeds than the bound fragments, are present in the plots.}\label{fig:density}
\end{figure}

Panel (a) of Figure~\ref{fig:density} shows each potentially observable fragment's position and velocity. The plot shows the high concentration of fragments at the galactic center as well as the wide velocity spread between bound and unbound fragments. An extremely small percentage of unbound fragments, 0.03\%, are still found within the galaxy; these are the few that are observed at a time near to their formation and ejection.

\begin{figure}[t]
\center
\includegraphics[width=\linewidth]{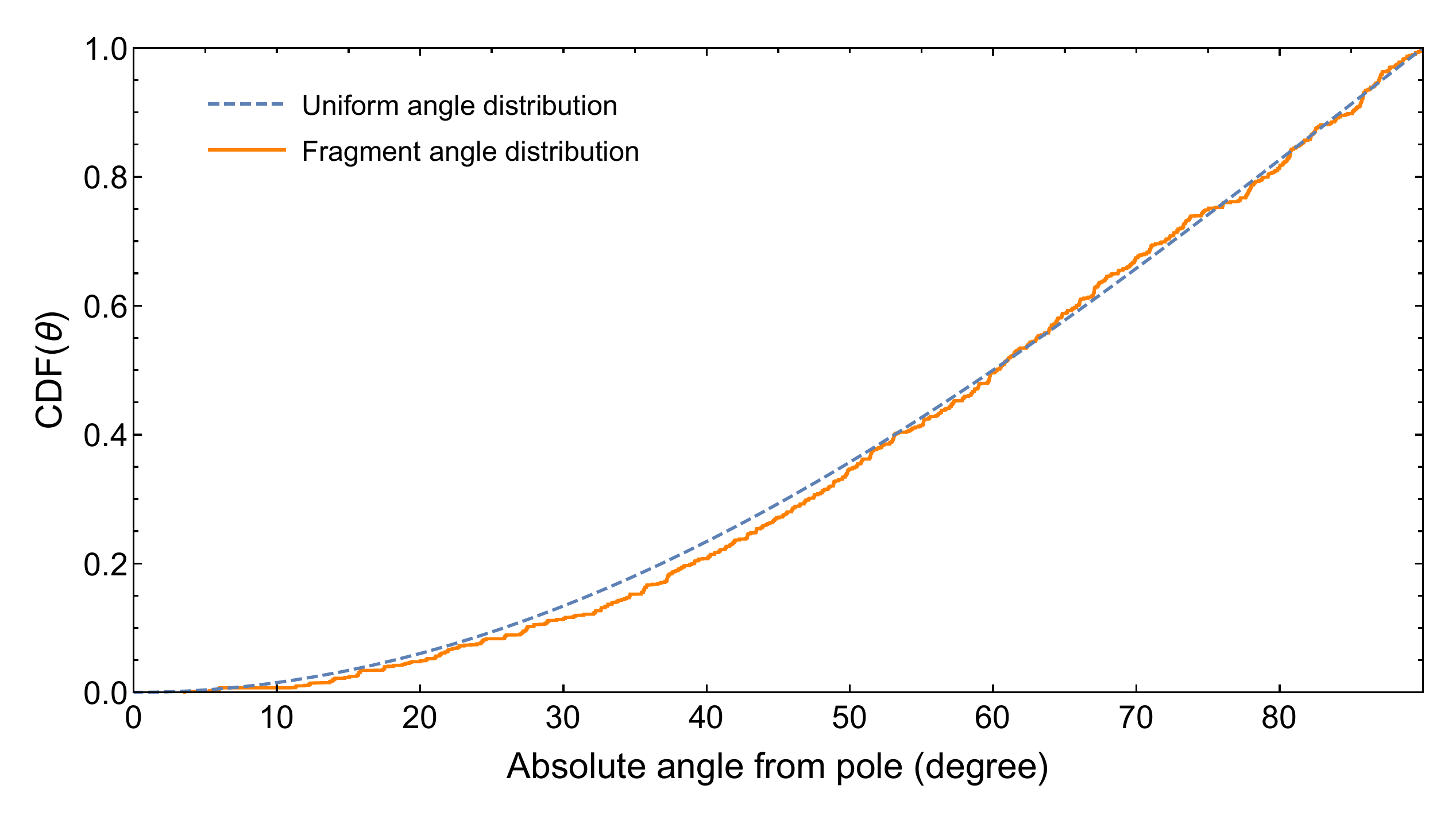}
\caption{Cumulative distribution function of azimuthal angle for potentially observable fragments as compared to a uniform angle distribution.}\label{fig:angles}
\end{figure}

Panel (b) of Figure~\ref{fig:density} shows the fragment positions in cylindrical coordinates, and reveals ``fingers of God''-like structures that reflect the mostly radial motion of the fragments within each stellar debris stream, as well as a noticeable deficit of fragments near the galactic poles. Because the velocity vectors are drawn uniformly in angle, one would assume that the angle at which the fragments are potentially observed would also be uniformly distributed. Figure~\ref{fig:angles} shows the cumulative density function of potentially observable fragment azimuths. We performed a Kolmogorov-Smirnov test to compare the distribution of these angles to a uniform angle distribution, and find a statistically significant difference between the two distributions, rejecting the null hypothesis that the fragments are distributed uniformly in angle with $p$-value = $2.5 \times 10^{-8}$. Examining the distribution, we find a noticeable deficit of fragments with azimuthal angle (measured from the pole) less than 45$^{\circ}$. This deficit suggests that the galaxy's disk component is able to bind some fragments that would otherwise be marginally unbound, increasing the fragment density in the disk plane.

To estimate a minimal distance at which a fragment might be observed near our sun we compute the local space density of fragments, which is related to how many bound fragments have been produced by tidal disruptions over the Milky Way's lifetime, $N_{\rm tot} = \Gamma_{\rm TDE} \tau_{\rm H}$. By counting the total fraction of fragments contained within a torus ${\cal N}_{\rm torus}$ with radius equal to the Sun's galactocentric distance of 8~kpc and tube radius 1~kpc, the spatial density of fragments near the Sun can be computed as
\begin{align}
n_{\rm frag} &= \frac{{\cal N}_{\rm torus}}{{\cal N}_{\rm frag}} \frac{\Gamma_{\rm TDE} \tau_{\rm H}}{N_{\rm TDE}}\\
&= 5.7~{\rm kpc}^{-3}\left(\frac{\Gamma_{\rm TDE}}{10^{-4}~{\rm yr}^{-1}}\right) \left(\frac{T_{\rm f}}{5 \times 10^{3}~{\rm K}}\right)^{9/4},\nonumber
\end{align}
meaning the typical distance to the nearest fragment is
\begin{align}
l_{\rm frag} &= 560~{\rm pc} \left(\frac{\Gamma_{\rm TDE}}{10^{-4}~{\rm yr}^{-1}}\right)^{-1/3}\nonumber\\
&\quad\times \left(\frac{T_{\rm f}}{5 \times 10^{3}~{\rm K}}\right)^{-3/4},
\end{align}
where we have propagated the temperature dependence of Equation (\ref{eq:mfrag}) into the above expressions.

\section{Discussion}\label{sec:discussion}

Given their potential proximity to our solar system, our work motivates questions regarding the fragments' observability. As each fragment forms, it possesses an initial temperature comparable to $T_{\rm f}$, briefly joins the Hayashi track, and then slowly radiates away its excess energy as it collapses.

The fragments produced in a tidal disruption event can be broadly thought of as approximately Jupiter-mass objects comprised of pure hydrogen and helium with a small fraction of the burning products of stellar evolution, with the exact composition depending on where exactly in the stream the fragment forms \citep[see e.g. Figure~10 of][]{Law-Smith:2017a}. With no nuclear reactions (the fragments are deuterium-free), there is no extra source of internal energy aside from gravity, and thus the fragments will be at their hottest at the onset of their collapse. This suggests that the fragments will resemble brown dwarfs with atypical compositions. Additional insight into the evolution of the material composition of tidally disrupted stellar fragments could be gained from utilizing stellar evolution simulation codes, such as {\tt MESA} \citep{Paxton:2011a}, to simulate their evolution more in depth.

As the fragments start their lives with temperatures on the order of several thousand K, their evolution is best described as a ``hot start'' \citep{Marley:2007a}, for which the luminosity evolves as \citep{Burrows:1993a,Marleau:2014a}
\begin{equation}
	L = 7.85 \times 10^{-6} L_{\odot}\left(\frac{M}{3 {\rm M}_{\rm J}}\right)^{2.641} \left(\frac{t}{10~{\rm Myr}}\right)^{-1.297}.\label{eq:lum}
\end{equation}

\begin{figure}[t]
	\centering
    \subfloat[Fragment mass distribution]{
    	\includegraphics[width=\linewidth]{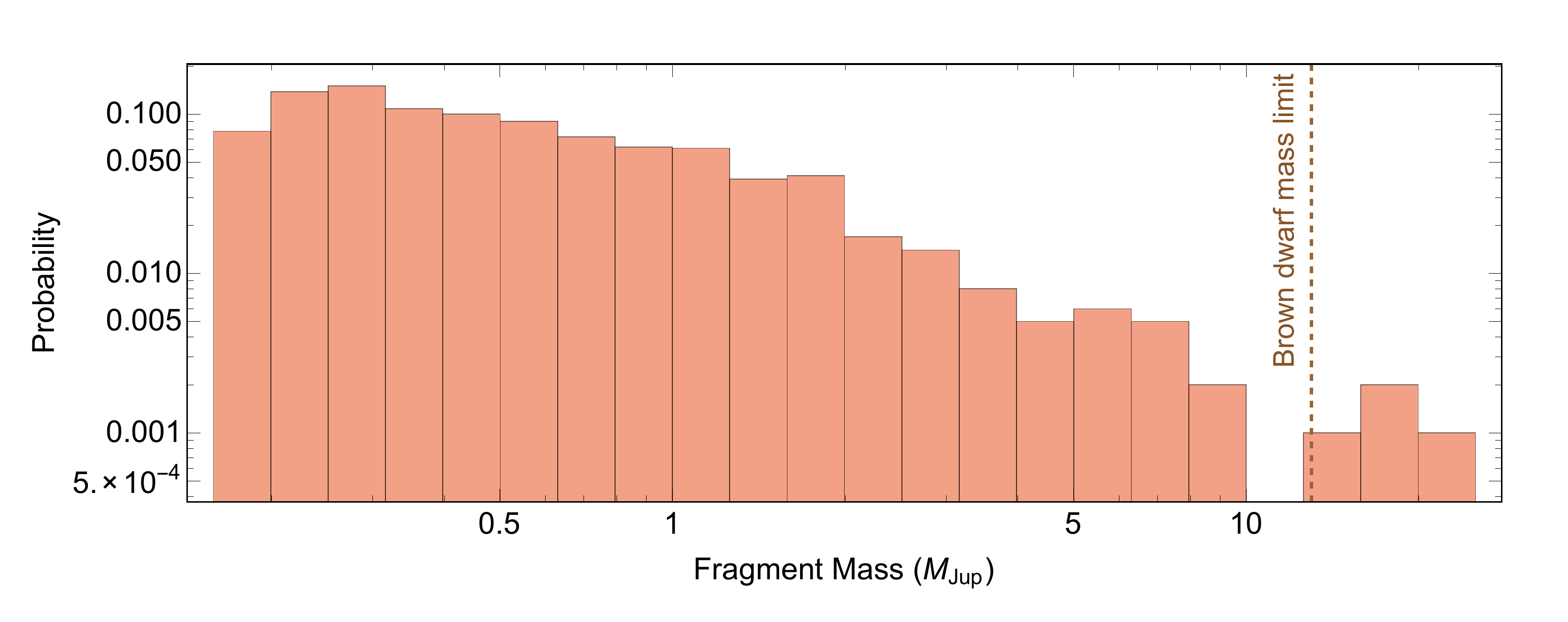}
    }\\*[3mm]
    \subfloat[Brightest apparent magnitude distribution]{
    	\includegraphics[width=\linewidth]{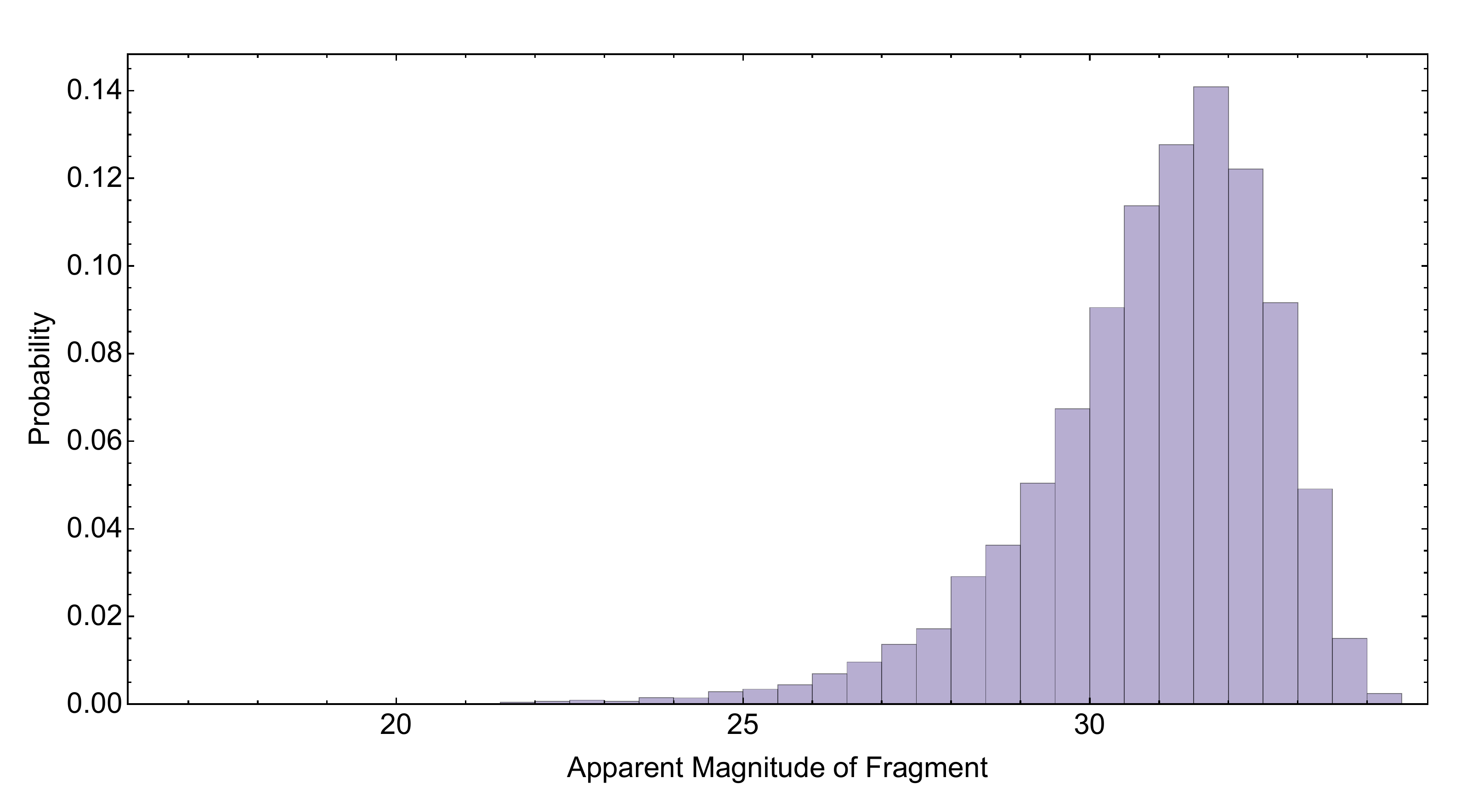}
    }
    \caption{Distributions of fragment mass and apparent magnitudes of the brightest fragments. For each simulated TDE, $M_{\rm frag}$ was calculated as outlined in Section \ref{sec:methods}. To calculate apparent magnitudes of fragments in the vicinity of the sun, we conducted $10^4$ random realizations by uniformly drawing an age for each fragment in the range [0, $10^{10}$~yr]. From these realiztations, we plot the minimum apparent magnitudes.}\label{fig:luminosity}
\end{figure}

In Panel (a) of Figure~\ref{fig:luminosity}, we show the distribution of fragment masses from our simulations. Most fragments have a mass somewhat less than that of Jupiter, but occasionally fragments with masses above the Brown dwarf limit form from the debris. These more massive fragments are significantly brighter, and are potentially more observable. Applying Equation~(\ref{eq:lum}) to these masses, and assuming a uniform fragment age distribution in the local neighborhood, we show in Panel (b) of Figure~\ref{fig:luminosity} the apparent bolometric magnitudes of the fragments with the highest apparent brightness that lies within 5~kpc of the Sun. The median apparent magnitude of the brightest fragment is 31.2, with a 10\% chance of the brightest fragment having magnitude $<28.6$, and a 1\% chance for a magnitude $<25.4$. The brightest fragment is typically found to lie within 1~kpc of the Sun, slightly further than the nearest fragment. Thus, while there is a small chance of a relatively young, massive fragment being discoverable with present-day facilities, it is most likely most fragments will only be found by future facilities with limiting survey magnitudes $\gtrsim 30$.

Our simulations have demonstrated that a majority of fragments produced by tidal disruptions in our galaxy are unbound to the SMBH and traveling at relativistic speeds -- the fastest of these being shot out at velocities $\sim 0.1 c$. Over 70\% of the unbound fragments travel a speed that allows them to reach the Milky Way even from the Virgo cluster, which lies at a distance of $\sim 20$~Mpc. Assuming an average galaxy density of $10^{-2}~{\rm Mpc}^{-3}$, the total number of extragalactic unbound fragments within 1~Mpc of the Milky Way is $8 \times 10^6$, comparable to the total number of bound fragments produced by Sgr~A*. The discovery of such fragments would provide additional compelling evidence that stars are regularly deposited onto lethal orbits about SMBHs where they are destroyed by tidal forces.

\bigskip
\acknowledgments

We thank John~A.~Johnson, Jorge Moreno, and the Banneker-Atzl\'{a}n Institute for their immense hospitality, support of our research, and continual affirmation that, as Benjamin Banneker so eloquently stated, \emph{the color of the skin is in no way connected with strength of the mind or intellectual powers}. This work utilized the {\tt Astropy} \citep{Astropy-Collaboration:2013a} and {\tt Scipy} \citep{Jones:2001a} \Python packages.

\bibliographystyle{yahapj}
\bibliography{library}

\end{document}